\documentclass[prl,amsmath,aps,twocolumn,superscriptaddress]{revtex4}
\usepackage{graphicx,xspace}
\usepackage{float}
\usepackage{amsmath}
\usepackage{amssymb}
\usepackage[normalem]{ulem}
\input{epsf}
\begin{document}
\newcommand{\scalar}[2]{\left \langle#1\ #2\right \rangle}
\newcommand{\me}{\mathrm{e}}
\newcommand{\mi}{\mathrm{i}}
\newcommand{\dif}{\mathrm{d}}
\newcommand{\period}{\text{per}}
\newcommand{\free}{\text{fr}}
\newcommand{\mq}[2]{\uwave{#1}\marginpar{#2}} 
\include{latexcommands}

\title{Increasing of entanglement entropy from pure to
random quantum critical chains.}
\author{Raoul Santachiara}
\email{santa@lpt1.u-strasbg.fr}
\affiliation{CNRS-Laboratoire Physique Theorique, Universit\'e Louis Pasteur\\
3 rue de l'Universit\'e, 67084 Strasbourg, C\'edex, France }
\begin{abstract}
It is known that the entropy of a block of spins of size $L$
embedded in an infinite pure critical spin chain diverges as the
logarithm of $L$ with a prefactor fixed by the central charge of the
corresponding conformal field theory. For a class of strongly random
spin chains, it has been shown that the correspondent block entropy
still remains universal and diverges logarithmically with an
"effective" central charge. By computing the entanglement entropy
for a family of models  which includes the $N$-states random Potts chain 
and the
  $Z_N$ clock model, we give some definitive answer to some recent
  conjectures about the behaviour of this effective central charge.
In particular,  we
  show that the ratio between the entanglement entropy in the pure and in the
  disordered system is model dependent and we  provide a series of
  critical models
  where the entanglement entropy  grows from the pure to the random case.
\end{abstract}
\maketitle

The effects of quenched randomness on quantum critical systems have
been the subject of an intense experimental and theoretical activity
for many years \cite{Bhatt}. Disordered one dimensional (1D) quantum models
provide an important framework to study these effects because the
quantum fluctuations are enhanced in this dimension and the
corresponding pure systems can be studied using a wide variety of
approaches, such as Bethe-ansatz solutions and conformal field theories
techniques.  For systems displaying strong spatial inhomogeneities,
many results can be obtained by the application of the strong
disorder renormalization group (SDRG) method \cite{Cecile}
introduced by Ma, Dasgupta and Hu \cite{Ma}. For a wide variety of
random quantum critical models, ranging from the random quantum
chains with discrete symmetries to the random Heisenberg models on
fractal lattices \cite{Fischer,
  fischer_heis,hyman,jolicoer,refael_2,latorre2,pierre,Satya,melin}, it has
been shown that the effective disorder grows indefinitely at larger
scales. This flow to strong disorder occurs in particular for the random spin
$1/2$-Heisenberg chain, the random $XX$ chain and the random
transverse field Ising model (RTFIM). The physics of these systems is
captured by the so called infinite random fixed point (IRFP). Some of
the main features of this point are a strong dynamical anisotropy and
an extreme broad distribution of physical quantities which manifests
through drastically different behaviour between average and typical
correlation functions.

Recently, Refael et Moore \cite{refael} have used the particular
properties at the IRFP of these models to compute the entanglement
of a block of spins of the corresponding chain.
 The entanglement of a segment of $L$ spins
with the remainder is defined as the von Neumann entropy of the
reduced density matrix $\hat{\rho}_L$ \cite{Delgado}:
\begin{equation}
\mathcal{S}(L)=-Tr \hat{\rho}_L \,\ln\, \hat{\rho}_L.
\label{def_entropia}
\end{equation}
In an infinite quantum critical chain without disorder, the
entanglement of the segment with the remaining sites is proportional
to $\ln L$ with a coefficient depending on the central charge $c$ of
the associated conformal field theory \cite{Holzhey,Vidal,pasquale}:
\begin{equation}
\mathcal{S}(L)\sim \frac{c}{3}\ln L
\label{entropia_pura}
\end{equation}
The main result of \cite{refael} is that the universal logarithmic
scaling of the entanglement entropy still holds at the IRFP of these
models, where the conformal invariance is lost. In particular they
showed that the entanglement entropy $\mathcal{S}_{RTFIM}$,
$\mathcal{S}_{RH}$ and $\mathcal{S}_{RXX}$ respectively for the RTFIM,
the random Heisenberg and $XX$ chain is:
\begin{eqnarray}
\mathcal{S}_{RH}(L)=\mathcal{S}_{RXX}(L)&\sim& \frac{\log 2}{3}\ln L
\nonumber\\
 \mathcal{S}_{RTIFM}(L)&\sim&\frac{\log 2}{6}\ln L
\label{entropia_rtfim}
\end{eqnarray}
Numerical evidences for the logarithmic scaling have been obtained
for the random Heisenberg and $XX$ chain in \cite{laflorencie,
pasquale2}.
  
Analogously to the role played by the central charge, the prefactor
of the logarithmic divergence can be a good estimator to characterise
the universality class of the random chains. The continuum limit of
the pure Heisenberg (H) and $XX$ chains (XX) are described by the
$c=1$ free boson theory (with different compactification radius) while
the pure quantum Ising (TFIM) is described by the $c=1/2$ free fermion
theory.  From (\ref{entropia_pura}), one has
$S_{H}(L)=S_{XX}(L)\sim\ln L /3$ and $S_{TFIM}(L)\sim\ln L /6$.
Hence, the ratio between the random and pure logarithmic prefactor is,
for these two models, the same.  This has suggested the intriguing
possibility that the coefficient appearing in the entropy scaling of
any random fixed point is the product of the central charge of the
associated pure critical system and an universal number ($\ln 2$).
The findings (\ref{entropia_rtfim}), from which the entanglement of
the pure systems turns out to be reduced by disorder, suggest also a
loss of entanglement along a general SDRG flow.  Note that the loss of
entanglement entropy was proposed in \cite{latorre} as a signal
of the irreversibility of the renormalization group (RG) flow . This
conjecture has been explored for the quantum Ising chain by varying
the transverse magnetic field and it was also found to be consistent
with the predictions for a Luttinger liquid with one single impurity
\cite{Levine}.

In this Letter we adopt the analysis of Refael and Moore to study the
entanglement entropy of a family of random quantum critical chains,
which includes the $N$-state random quantum Potts chain and the $Z_N$
clock model. By means of the SDRG procedure, we show that all these
random chains are attracted by the infinite random fixed point of the
RTFIM.  Computing the associated entanglement entropy, we show that
that the correction from a pure conformal fixed point to the random
fixed point depends on $N$ and thus is model-dependent. Moreover we
find that for the random parafermionic $Z_N$ chains (defined later)
with $N>41$, the entanglement is greater than the one in the pure
system, thus providing a counter-example to the above conjecture.

We consider a chain of spins with $N$ states
$|s_i\rangle=|0_i\rangle,|1_i\rangle,\cdots,|(N-1)_i\rangle$
where $i$ indicates the lattice site. The random quantum spin
model we study is defined by the Hamiltonian
$\mathcal{H}_N$:
\begin{eqnarray}
 \mathcal{H}_N=&& -\sum_i J_{i,i+1} \sum_{n=1}^{N-1}
 \alpha_n \left(\bar{S}^{z}_i S^{z}_{i+1}\right)^n-\nonumber \\&&
-\sum_i h_i \sum_{n=1}^{N-1} \alpha_n \Gamma_{i}^n.
\label{e:def_model}
\end{eqnarray}
in terms of the operators $S^{z}=e^{2i\pi/N q}\delta_{q,q'}$ (with
$q,q'=0 \cdots N-1$), their hermitian conjugates $\bar{S}^{z}$ and
spin raising operators $\Gamma|s\rangle=|s+1,\, \mbox{mod}\,
N\rangle$.  The couplings $J_{i,i+1}$ and the transverse fields
$h_{i}$ are independent positive random variables drawn from some distribution.
The coefficients $\alpha_n$ satisfy $\alpha_n=\alpha_{N-n}$ to assure
the Hamiltonian to be hermitian and they are assumed to be disorder
independent.  For each set of values $\{\alpha_n\}$, it exists a
transformation
from site variables to bond variables $
\bar{S}^{z}_i S^{z}_{i+1}=\Gamma^{*}_i; \quad \prod_{j\leq
i}\Gamma_j=S^{z,*}_i$ 
which yields the same Hamiltonian with the $J$'s and $h$'s
interchanged. The model (\ref{e:def_model}) is thus provided of a
duality transformation. In the following we will always assume the
couplings and fields to have initial equal distribution.

If all the coefficients $\alpha_{n}$ are equal to unity,
$\alpha_{n}=1$ for $n=1\cdots N-1$, the Hamiltonian
(\ref{e:def_model}) has a permutational symmetry $S_N$ and it
defines the random quantum $N$-states Potts model.  The case
$\alpha_{n}=\delta_{1,n}$ corresponds to the random quantum $Z_N$
clock models.  Senthil and Majumdar \cite{Satya} showed that these
two models, with ferromagnetic couplings, are attracted to the IRFP
of the RTFIM.  As they noticed, this implies that the scaling
behaviour of many physical quantities (such as the magnetisation or
the mean correlation functions) depend on statistical properties
entirely given by the IRFP distributions which are $N$-independent.
So, despite the fact that in the absence of disorder these systems
have different type of correlations, the low-energy  behaviour of
the random systems presents $N$-independent scaling functions and
exponents. On the other hand, the entanglement entropy at the IRFP
of these models is, as we show later, sensitive to the number of
spin states. This quantity can then discriminate
between models with different $N$.

Here we focus also on another class of random quantum spin chains
defined by the Hamiltonian (\ref{e:def_model}) with
$\alpha_{n}=\sin(\pi/N)/\sin(\pi n/N)$.  The interest in this model is
motivated by the following facts: i) the corresponding pure model is
critical at the self dual line $J=h$ \cite{Alcaraz} and its
fluctuations are governed by the $Z_N$ parafermionic field theories
with central charge $c_N=2(N-1)/(N+2)$ introduced in \cite{Fateev}
(the case $N=2$ and $N=3$ correspond to the Ising and the $N=3$ state
Potts model respectively). ii) The SDRG procedure can be performed for
all this family of theories. iii) The scaling of entanglement entropy
for the $Z_N$ spin chain can be computed using the properties of the
IRFP and compared with its value at the parafermionic conformal
critical points.  Henceforth we refer to this model as the $Z_N$
parafermionic spin chain.

We apply the SDRG approach to the model (\ref{e:def_model}) by
assuming the coefficients $\alpha_{n}\leq 1$. The renormalization
equations are obtained by successively eliminating the maximum of the
amplitudes of the bonds $J_{i,i+1}$ and fields $h_{i}$, getting an
effective Hamiltonian for low energy degrees of freedom. At each
decimation step, if this maximum turns out to be the field $h_i$, the
corresponding spin $|s_i\rangle$ is frozen in the state
$|s_i\rangle=1/\sqrt{N}(|0_i\rangle+|1_i\rangle+\cdots+|(N-1)_i\rangle)$ which
is the ground state of the dominant term $-h_i \sum_{n=1}^{N-1}
\alpha_n \Gamma_{i}^n$. This generates an effective new coupling
$\tilde{J}_{i-1,i+1}$ between adjacent sites. The coefficients
$\alpha_{n}$ enter into the renormalization as well. Degenerate second
order perturbation theory yields:
\begin{equation}
 \tilde{J}_{i-1,i+1}=\frac{J_{i-1}J_{i+1}}{\kappa_{1}h_{i}}\quad \quad
\tilde{\alpha}_{n}=\alpha_{n}^2\frac{\kappa_{1}}{\kappa_{n}}.
\label{e:jalpha_flow}
\end{equation}
where the $\tilde{\alpha}_{n}$ are the renormalized coefficients and
\begin{equation}
 \kappa_{n}=\frac{1}{2}\sum_{m=1}^{N-1} \alpha_{m}\left[1-\cos\left(\frac{2\pi m n}{N}\right)\right].
\label{e:k_def}
\end{equation}
On the other hand, if the maximum is the coupling $J_{i,i+1}$, we set
spin $|s_i\rangle$ and spin $|s_{i+1}\rangle$ to have the same value,
$|s_i\rangle=|s_{i+1}\rangle$. The effective field
$\tilde{h}_{i}$ acting on the ferromagnetic cluster $|s_i\rangle \otimes |s_{i}\rangle$ is:
\begin{equation}
 \tilde{h}_{i}=\frac{h_{i}h_{i+1}}{\kappa_{1}J_{i,i+1}},
\end{equation}
as it could be guessed by duality.

\begin{figure}
\centerline{
\epsfxsize=6.5cm
\epsfclipon
\epsfbox{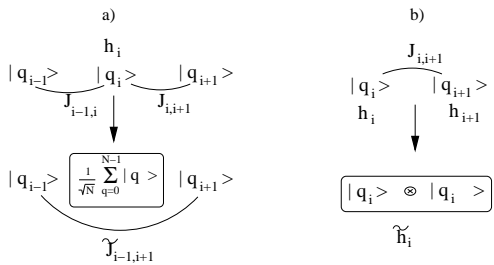}}
\caption{Schematic picture of the two kind of decimation step: a) Field decimation. b) Coupling decimation.}
\label{decimation}
\end{figure}

Due to our initial assumptions on the coefficients $\{\alpha_n\}$, we
have $\kappa_1\geq 1$ for $N\geq 2$.  This assures that, during the
renormalization procedure, the energy scale is  decreasing:
the fixed point of the problem is thus expected to be strongly
attractive, even for weak disorder.  It is important also to
remark that the disorder drives the system away from the parafermionic
critical point but the self-duality is preserved during
the decimation procedure.

Starting with the Potts values $\alpha_{n}=1$ for $n=\cdots N-1$, the
coefficients do not renormalize during the decimation procedure. In
other words the permutational symmetry $S_N$ is not broken along the
flow of the couplings.  On the contrary, by applying iteratively Eq.\ref{e:jalpha_flow} and Eq. \ref{e:k_def}, the coefficients
$\{\tilde{\alpha}_{n}\}$ defining the $Z_N$ parafermionic spin chain
renormalize quickly to the values $\tilde{\alpha}_{n}\to\delta_{1,n}$.
We can thus conclude that the random parafermionic $Z_N$ spin
chain and the random $Z_N$ clock model share the same low-energy
behaviour. In particular the couplings and fields distribution
is attracted towards the IRFP
distribution.

\begin{figure}
\centerline{
\epsfxsize=6.5cm
\epsfclipon
\epsfbox{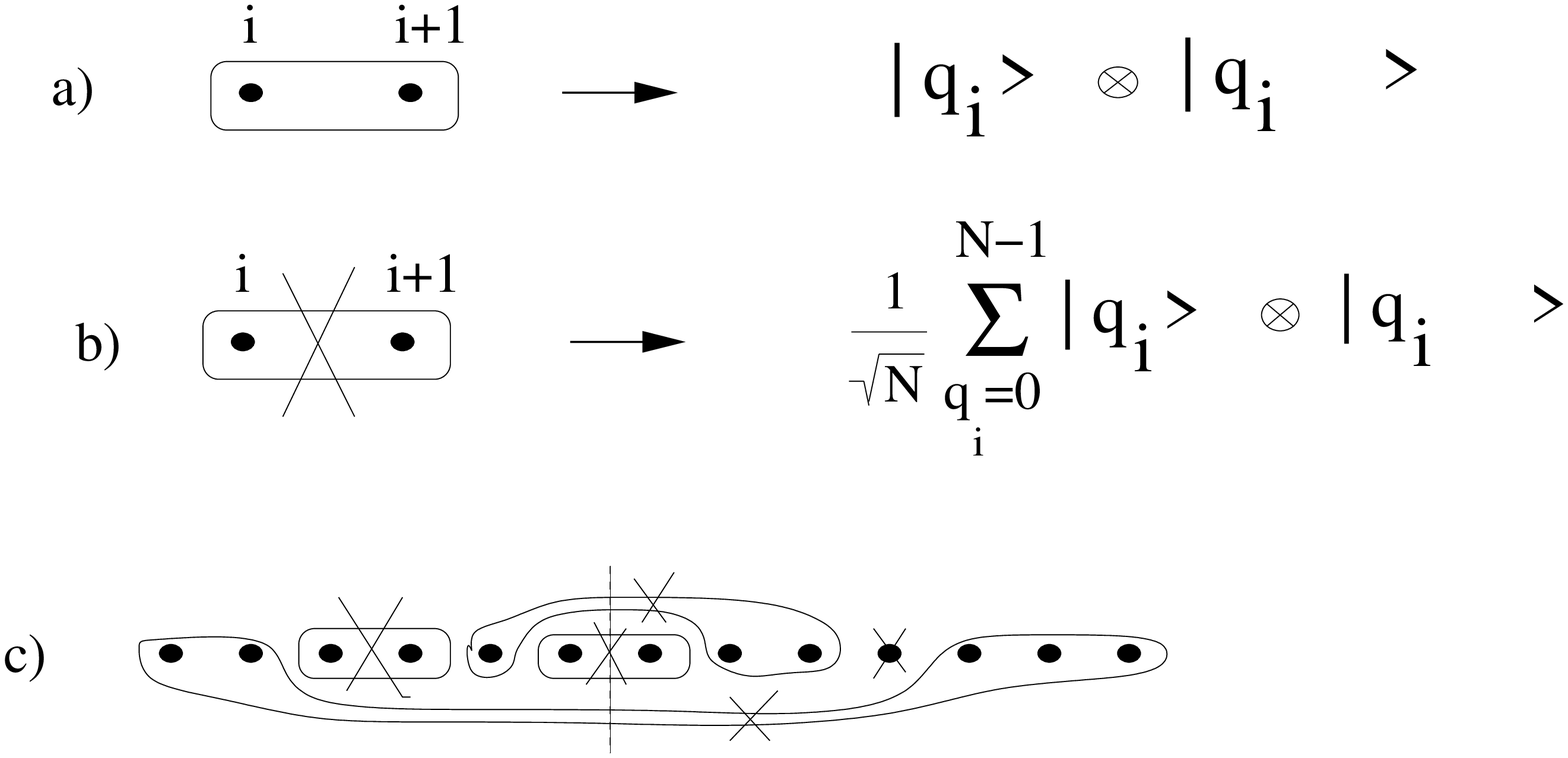}}
\caption{a) The state describing two spins decimated by the
couplings. b) The state of a cluster of two spins after the field
decimation. c) The contribution of the entanglement is given by the
field decimated cluster which cross an edge of the segment (dashed
line). In this example, we have three of such clusters.}
\label{decimation2}
\end{figure}

The analysis of Refael and Moore can then be directly generalised to
the more general random quantum Hamiltonian (\ref{e:def_model}).  The
critical fixed point distribution yields a concentration $n_{\Gamma}$
of cluster at the log-energy scale $\Gamma$ which scales as
$n_{\Gamma}\sim 1/\Gamma^2$.  Hence the ground states of all these
models present clusters which form over an average length $\lambda
\sim \Gamma^2$.  Let us consider in detail a cluster of $s$ spins. We
want to compute  the entanglement entropy of a subsystem of $r$ spins with
the remaining $s-r$ spins of the cluster.  If the cluster has
been decimated by the field, it is frozen in the state $|\psi\rangle$:
\begin{equation}
|\psi\rangle=
\frac{1}{\sqrt{N}}
\sum_{q=0}^{N-1}
\overbrace{|q\rangle
\otimes|q\rangle\otimes\cdots\otimes|q\rangle}^{\mbox{s
times}}.
\label{purestate}
\end{equation}
The entries of the reduced density matrix $\hat{\rho}_r$ can be
defined in the spin basis $\phi^{i}_r$, $i,j=1..2^r$ as:
\begin{equation}
\langle \phi^{i}_r|\rho_r|\phi^{j}_r\rangle=
\sum_{k,l=1}^{2^{s-r}}\langle\phi^{i}_r| \otimes \langle \phi^{k}_{s-r}|
 \psi\rangle \langle \psi|\phi^{l}_{s-r}\rangle \otimes |\phi^{j}_{r}\rangle,
\label{def_density}
\end{equation}
where one traces over the basis $\phi^{i}_{s-r}$ of the other $s-r$
spins of the cluster. Using the wave function (\ref{purestate}) in
(\ref{def_density}), one gets that the entanglement of a subsystem of
$r$ spins with the remaining $s-r$ ones of a frozen cluster is $\log
N$ for each $r$ and $s$. Returning to the entanglement of a segment
$L$ embedded in an infinite chain, the frozen clusters which are
totally inside or outside the segment $L$ do not affect the total
entanglement. The only contribution comes from the number of decimated
clusters which cross the two edges, each of such clusters adding $\log
N$ to the total value of the entanglement (see Fig.\ref{decimation2}).
One has then
\begin{equation}
\mathcal{S}^{dis}_N(L)\sim 2\, \log N\, p\,\langle D \rangle_L
\end{equation}
where $\mathcal{S}^{dis}_N(L)$ is the entanglement of a segment $L$ in
the disordered quantum chain (\ref{e:def_model}), $\langle D
\rangle_L$ is the average number of decimation which occur over an
edge of the segment $L$, $p$ is the probability to have a field
decimation instead of a coupling one. Due to the self-duality
of these models, preserved along the SDRG, one has
$p=1/2$.  The value of $\langle D \rangle_L$ does not depend on $N$:
it is entirely given from the fixed point distribution and it has been
computed in \cite{refael} where it was found $\langle D \rangle_L=1/3
\log L $.  Collecting all these results we finally arrive at the value
of entanglement for the family of random models (\ref{e:def_model})
with $\alpha_{n}\leq 1$ :
\begin{equation}
\mathcal{S}^{dis}_N(L)\sim\frac{\log N}{6}\log L.
\end{equation}

The entanglement $\mathcal{S}^{dis}_N(L)$ depends thus on the number
of states $N$. Hence, differently from other quantities which are
$N$-independent, it represents a good estimator to characterise the
universality class of the random chains.  Note that this quantity
alone is not sufficient to identify a random phase transitions as it
can be seen, for instance, from the fact that the random quantum Potts
model with symmetry $S_N$ have the same entanglement of the models
with a lower symmetry $Z_N$. This is analogous to what happens in
conformal field theory where different models can have the same
central charge.

Taking into account the value of the central charge $c_N$, the
entanglement $\mathcal{S}^{pure}_N(L)$ of the pure $Z_N$
parafermionic spin chain is
\begin{equation}
\mathcal{S}^{pure}_N(L)\sim\frac{2(N-1)}{3(N+2)}\ln L.
\end{equation}
Comparing the values $\mathcal{S}^{pure}_N$ and
$\mathcal{S}^{dis}_N(L)$, we can rule out the existence of a
model-independent correction which links the values of the
entanglement of the pure and of the disordered system. Actually, there
is no any clear relationship between the value of the entanglement
prefactor in the disordered chain and the central charge associated to
the pure chain. It would be interesting to check other observables,
for instance the scaling of the free energy at the IRFP, which could
clarify better if the value of this prefactor behaves as an
``effective central charge'', as recently proposed
\cite{refael,laflorencie}.

It is interesting to note that :
\begin{equation}
\mathcal{S}^{dis}_N(L)> \mathcal{S}^{pure}_N(L)\quad \mbox{for} N>41,
\end{equation}
thus invalidating the recent conjecture of a loss of entropy from
pure to disordered critical chain.  This conjecture followed along
the lines of the c-theorem \cite{ctheorem} which states the
existence of a function of the couplings monotonically decreasing
along the renormalization group. For this reason it is interesting
to mention the fact that, in the presence of disorder, the unitarity
(one of the key hypotheses of the $c$-theorem) is in general lost.
This is known to happen, for instance, in the case of the
two-dimensional random bond $3$-states Potts model which flows
towards a random fixed point where the conformal symmetry is
restored \cite{Dotsi_picco_pujol,picco_jesper}. In this case, one
can show that the energy-energy operator, which determines the
properties of the RG equations, gets a negative norm in the disorder
limit. This explains the fact that, for this model, the central
charge associated to the disordered random fixed point is greater
than the one associated to the pure fixed point. On the other hand,
the loss of unitarity does not necessarily imply the increasing of
the central charge, as it has been explicitly shown for instance in
\cite{raoul_wdn}, where a particular class of disordered conformal
field theories was studied.  In an analogous way, we think that, in
the presence of disorder, one cannot predict the behaviour of the
entanglement on the basis of a generalised $c-$theorem, as our
results clearly show.

To conclude, in this Letter we have shown that the entanglement entropy of a class
of random quantum critical chains, including the random quantum Potts
and the random quantum clock chain, obey an universal logarithmic
scaling.  We have confirmed thus the entanglement as an important
quantity to characterise (although not completely) the universality
class of this entire series of disordered critical models.  Moreover
our results provide a definitive answer to some recent hypotheses on
the entanglement entropy in the random quantum critical chains.  By
considering the $Z_N$ parafermionic spin chains, we could compare the
value of the central charge of the pure system with the value of the
entanglement entropy of the disordered one for each value of $N$. We
found that the ratio between the central charge and the prefactor of
the entanglement logarithmic scaling of the disordered models depends
on $N$. We could thus exclude the existence of a model-independent
correction between these two quantities. Moreover, we provide some
explicit models where the entanglement of the pure system is smaller
than the entanglement of the disordered one. Thus the recent
conjecture about a generalised $c$-theorem concerning the entanglement
does not hold when the disorder is present.

I would like to thank Daniel Cabra, Mirko
Cinchetti, Vladimir Dotsenko, Andreas Honecker, Nicolas Laflorencie,
Pierre Pujol and Alberto Rosso for stimulating and interesting
discussions. I also thank the CNRS for financial support.


\begin{thebibliography}{10}
\bibitem{Bhatt} For a review, see: R.~N.~ Bhatt, in {\it Spin Glasses
    and Random Fields}, edited by A.~P.~Young (World Scientific,
  Singapore, 1998) p. 225.

\bibitem{Cecile} For a review, see: F.~Igl\'oi and C.~
  Monthus, cond-mat/0502448.

\bibitem{Ma} S.~K.~Ma, C.~Dasgupta and C.~K.~Hu, Phys. Rev. Lett. {\bf
    43}, 1434 (1979);  C.~Dasgupta and S.~K.~Ma,  Phys. Rev. B {\bf
22}, 1305 (1980).

\bibitem{Fischer}
D.~S.~Fisher, Phys. Rev. B {\bf 22}, 1305 (1980), Phys. Rev. Lett.
{\bf 69}, 534 (1992);

\bibitem{fischer_heis}
D.~S.~Fisher, Phys. Rev. B. {\bf 50}, 3799 (1994)

\bibitem{hyman}
R.~A.~Hyman and K.~Yang, Phys. Rev. Lett. {\bf 78}, 1783 (1997).

\bibitem{jolicoer} C.~Monthus, O.~Golinelli and Th.~Jolicoer, Phys.
  Rev. Lett. {\bf 79},  3254 (1997).

\bibitem{refael_2} G.~Refael, S.~Kehrein and D.~S.~Fisher, Phys.
  Rev. B. {\bf 66},  060402 (2002).


\bibitem{latorre2}
J.~.I.~Latorre, R.~Orus, E.~Rico, and J.~Vidal, 
Phys.Rev. A {\bf 71} 064101 (2005).

\bibitem{pierre}
D.~Carpentier, P.~Pujol, and K.~U.~Giering, 
cond-mat/0506597.

\bibitem{Satya}
T.~Senthil and S.~N.~ Majumdar,
\newblock Phys. Rev. Lett. {\bf 76}, 3001 (1996)

\bibitem{melin}
R.~Melin, B.~Doucot and F.~Igloi,
\newblock Phys. Rev. B. {\bf 72}, 024205 (2005)

\bibitem{refael}
G.~Refael and J.~E.~Moore,  Phys. Rev. Lett. {\bf 93}, 260602 (2004).

\bibitem{Delgado}
A.~Galindo, M.~A.~Martin-Delgado
Rev.Mod.Phys. 74 (2002) 347-423; quant-ph/0112105

\bibitem{Holzhey}
C.~Holzjey, F.~Larsen and F.~Wilczek,
\newblock Nucl. Phys. B {\bf 424}, 443 (1994)

 \bibitem{Vidal}
G.~Vidal, J.~I.~Latorre, E.~Rico and A.~Kitaev, 
\newblock Phys. Rev. Lett. {\bf 90}, 227902 (2003)

\bibitem{pasquale}
 P.~Calabrese and J.~Cardy, 
\newblock J. Stat. Mech. {\bf 06}, 002 (2004).

\bibitem{laflorencie}
N.~Laflorencie,  Phys. Rev. B. {\bf 72},  R140408 (2005).

\bibitem{pasquale2}
G.~De~Chiara, S.~Montenegro, P.~Calabrese and R.~Fazio,
 cond-mat/0512586 .

\bibitem{latorre}
J.~.I.~Latorre, C.~A.~Lutken, E.~Rico, and G.~Vidal, 
Phys.Rev. A {\bf 71} 034301 (2005).

\bibitem{Levine}
G.~Levine,  Phys. Rev. Lett. {\bf 94},  062301 (2005).



\bibitem{Fateev}
V.~A.~Fateev and A.~B.~Zamolodchikov, 
\newblock Sov. Phys. JETP {\bf 62}, 215 (1985)

\bibitem{Alcaraz}
F.~C.~Alcaraz, 
\newblock J.Phys.A:Math.Gen. {\bf 20}, L623 (1987).

\bibitem{ctheorem}
A.~Zamolodchikov, JETP~Lett. {\bf 43}, 730 (1986).




\bibitem{Dotsi_picco_pujol}
Vl.~Dotsenko, M.~Picco and  P.~Pujol,
\newblock Nucl. Phys. {\bf B455}, 701 (1995).

\bibitem{picco_jesper}
J.~L.~Jacobsen and M.~Picco, 
Phys. Rev E {\bf 61} R13 (2000)

\bibitem{raoul_wdn}
Vl.~Dotsenko, X.S. Nguyen, and
R.~Santachiara, 
\newblock Nucl. Phys. B {\bf 613}, 445-471 (2001).


\end{thebibliography}
\end{document}